\documentclass[12pt]{l4dc2023}
\usepackage{times}
\usepackage{mathrsfs}

\usepackage{bm}
\usepackage{color}
\usepackage{lipsum}
\usepackage{algorithm}
\usepackage{algorithmic}
\usepackage{wrapfig}
\usepackage{graphicx}
\usepackage{comment}
\usepackage{IEEEtrantools}
\usepackage[capitalise]{cleveref}
\usepackage[font={small}]{caption}
\usepackage{hyperref}

\DeclareMathOperator{\Equaldef}{\overset{def}{=}}

\newtheorem{assumption}{Assumption}

\newcommand{\ip}[2]{\left\langle#1,#2\right\rangle}
\newcommand{\norm}[1]{\left\lVert#1\right\rVert}
\renewcommand{\(}{\left(}
\renewcommand{\)}{\right)}
\renewcommand{\[}{\left[}
\renewcommand{\]}{\right]}

\def \E {\mathbf{E}}

\def \< {\langle}
\def \> {\rangle}

\title[Networked top-$k$ Selection]{Top-$k$ data selection via distributed sample quantile inference} 

\author{%
 \Name{Xu Zhang} \Email{xuzhang\_cas@lsec.cc.ac.cn}\\
 \addr 
 Chinese Academy of Sciences
 \AND
 \Name{Marcos M. Vasconcelos} \Email{marcosv@vt.edu}\\
 \addr 
Virginia Polytechnic Institute and State University
}

\begin{document}

\maketitle

\begin{abstract}%
We consider the problem of determining the top-$k$ largest measurements from a dataset distributed among a network of $n$ agents with noisy communication links. We show that this scenario can be cast as a distributed convex optimization problem called sample quantile inference, which we solve using a two-time-scale stochastic approximation algorithm. Herein, we prove the algorithm's convergence in the almost sure sense to an optimal solution. Moreover, our algorithm handles noise and empirically converges to the correct answer within a small number of iterations.
\end{abstract}


\section{Introduction}

The selection of the $k$ largest quantities among a set of $n$ real numbers is of great importance in many applications of interest including machine learning \citep{rejwan2020top}, data mining \citep{tang2017extracting}, and information retrieval \citep{shi2019distributed}. These are known as \textit{top}-$k$ strategies.  Top-$k$ strategies also play a role in remote sensing \citep{Zhang:2022} and selection of the most informative neighbors for distributed optimization \citep{Verma:2021}. While this problem is trivial in the centralized case, where all the data is available at a single agent (server), it is a non-trivial problem in applications where the data is distributed over a network of clients. Here, we study the design of a distributed top-$k$ strategy over a network of many agents. 

\begin{wrapfigure}{r}{0.4\textwidth}
\vspace*{-0.6in}
 \centering
 \includegraphics[width=0.4\columnwidth]{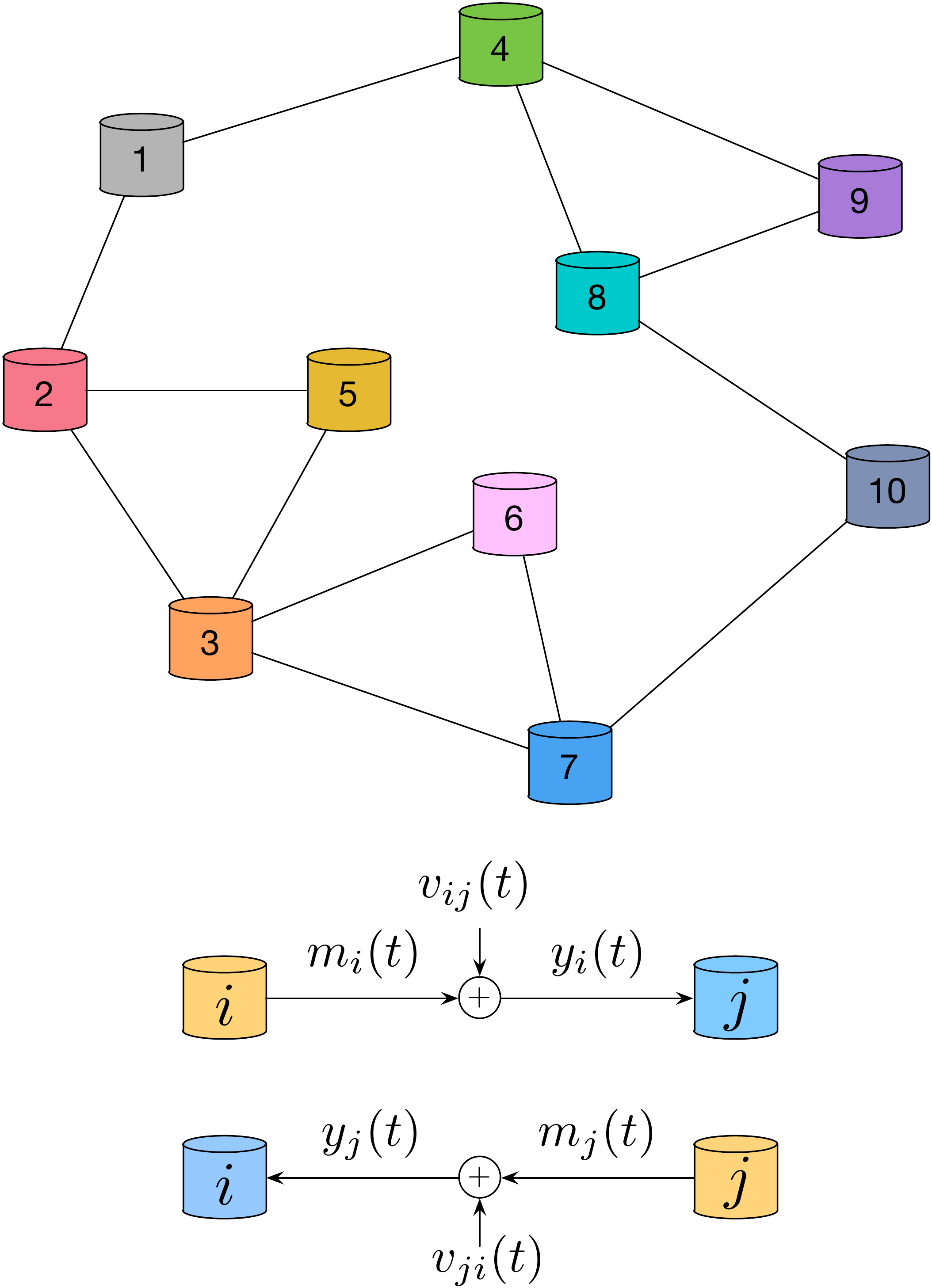}
\caption{System model -- A dataset is distributed accross multiple clients. Two clients can communicate if there is a link in between them. The communication links are noisy.} 
\label{fig:system}
\end{wrapfigure}

Consider a decentralized system in \cref{fig:system} where the agents can only communicate with their neighbors over a noisy channels. Each agent has its own measurement and our goal is to select the $k$ largest ones by learning an optimal threshold. Once this threshold is properly computed, each agent independently declares whether its measurement is above the threshold, and therefore that it has one of the top-$k$ measurements. One possible strategy is to relate the threshold with a sample quantile, which can be cast as quantile inference problem. Remarkably, this problem admits a natural decomposition as a distributed nonsmooth convex optimization problem. Herein, we provide an algorithm based on a two-time-scale subgradient method and perform its corresponding convergence analysis.

There exists a significant literature on the design of top-$k$ strategies in different settings. To the best of our knowledge 
\citep{babcock2003distributed}
proposed the first distributed algorithm to find the $k$ largest values in a federated setting using a technique called range caching. 
\cite{ustebay2011efficient} provided a top-$k$  algorithm to compute the top-$k$ entries based on selective gossip. 
 \cite{rajagopal2006universal} studied the distributed estimation of an arbitrary quantile in a communication-constrained federated scenario. Compared with the above papers, \cite{haeupler2018optimal} proposed faster optimal gossip algorithms for quantile computation by designing two types of tournament mechanisms. However, their algorithms are very sensitive to communication noise and do not converge to the exact quantile over noisy channels.

 Our work is closely related to the work of \cite{lee2018consensus}, where a two-time-scale algorithm to estimate a sample quantile in a decentralized way is claimed to converge in the mean squared sense. However, due to a technicality, the proof of convergence in \citep{lee2018consensus} fails to hold due to the lack of monotonicity of the sequence $\|\mathbf{E}[\bar{w}(t)] - \theta_p \mathbf{1}\|$. We avoid this technical difficulty by considering the convergence in the almost sure sense, combining the classical result of \cite{robbins1971convergence} with a modern result on nonsmooth optimization by \cite{yi2021distributed}.


\section{Problem setup}

We begin our analysis by relating the problem of computing the top-$k$ observations with quantile inference, which is a convex optimization problem. Consider a collection of $n$ agents, $[n]=\{1,\cdots,n\}$, interconnected by a time-invariant, undirected, communication graph $\mathcal{G}=([n],\mathcal{E})$. Each agent holds a non-negative real number, which may represent a sensor measurement, a belief, or an opinion. Throughout this paper, we will simply refer to them as \textit{data}. Let $z_i \in \mathbb{R}$ be the data of the $i$-th agent. The goal of the team of agents is to determine in a distributed fashion the $k$ agents holding the top-$k$ largest data. We want to do this efficiently. Moreover, the communication among agents occurs in the presence of additive noise.

At first, one may be inclined to consider the following strategy: Each agent keeps a list of  $k$ entries in its memory. At time $t$ each agent sends this list to its neighbors. At time $t+1$, every agent updates its list with by selecting the top-$k$ received data and discarding the rest. Each agent sorts its list and repeats. While this simple scheme converges to the top-$k$ results in finite time, it has two main drawbacks. First, this scheme requires noiseless communication channels of $k$ real numbers per channel use. Even the slightest amount of noise will cause the algorithm to diverge. Second, it requires a memory with size $k$. If $k\sim \mathcal{O}(n)$, the communication and storage requirements will quickly turn the problem of finding the top-$k$ observations accross the network prohibitive. 

On the other hand, this problem can be conveniently cast into the framework of distributed convex optimization, and admits an implementation where a single real number is exchanged and a single unit of memory is updated at each time. Furthermore, this algorithm is robust to the presence of noise. Consider the problem of infering the sample quantile from the dataset containing all of the agents' individual data points $\mathcal{D}\Equaldef\{z_i\}_{i=1}^n$. Let $\widehat{F}(\xi ; \mathcal{D})$ denote the empirical cumulative distribution function of the data set $\mathcal{D}$, defined as:
\begin{equation}
    \widehat{F}(\xi ; \mathcal{D})\Equaldef \frac{1}{n} \sum_{i=1}^{n} \mathbf{1}( z_i \leq \xi).
    \end{equation}
Let $p\in(0,1)$. The (sample) $p$-quantile is defined as
    \begin{equation}\label{eq:quantile}	\theta_{p}\Equaldef \inf \Big\{\xi \ \Big| \ \widehat{F}(\xi ; \mathcal{D}) \geq p \Big\},
	\end{equation}

    \begin{figure}
 \centering
 \includegraphics[width=0.85\columnwidth]{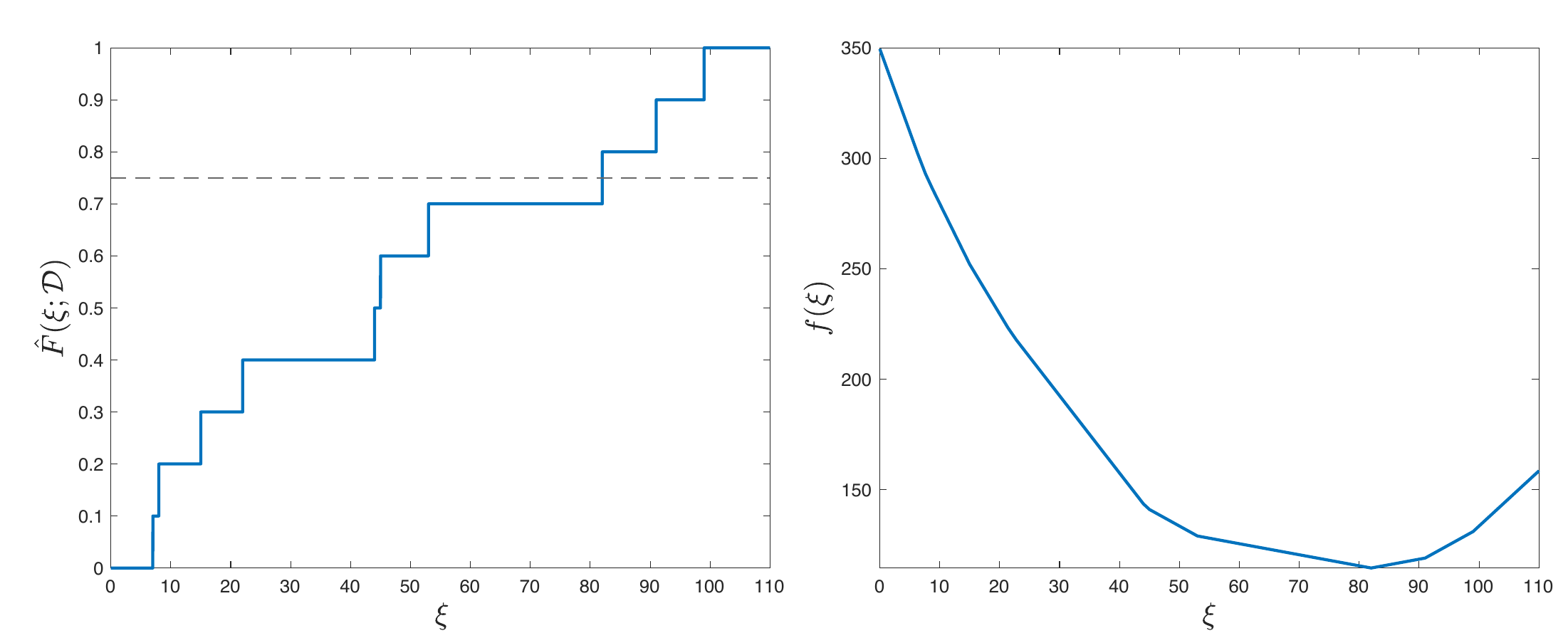}
 \caption{Empirical CDF and aggregate score function for $k=3$.} 
\label{fig:ECDF}
\end{figure}

A classic result in quantile regression relates \cref{eq:quantile} to the solution of the following optimization problem \cite[Section 1.3, pp. 7--9]{Koenker:2005}: \begin{equation} \label{eq:quantile_cvx}
    	\theta_p = \arg \min_{\xi\in \mathbb{R}} \sum_{i=1}^n \rho_p \big(z_i-\xi \big),
   \ \ \text{where} \ \
    	\rho_p(x) \Equaldef \begin{cases}
    		(p-1)x & \text{if} \ \ x < 0 \\
    		p x & \text{if} \ \ x \geq 0 . \\
    	\end{cases}
    \end{equation}

\begin{wrapfigure}{l}{0.4\textwidth}
 \centering
 \includegraphics[width=0.3\columnwidth]{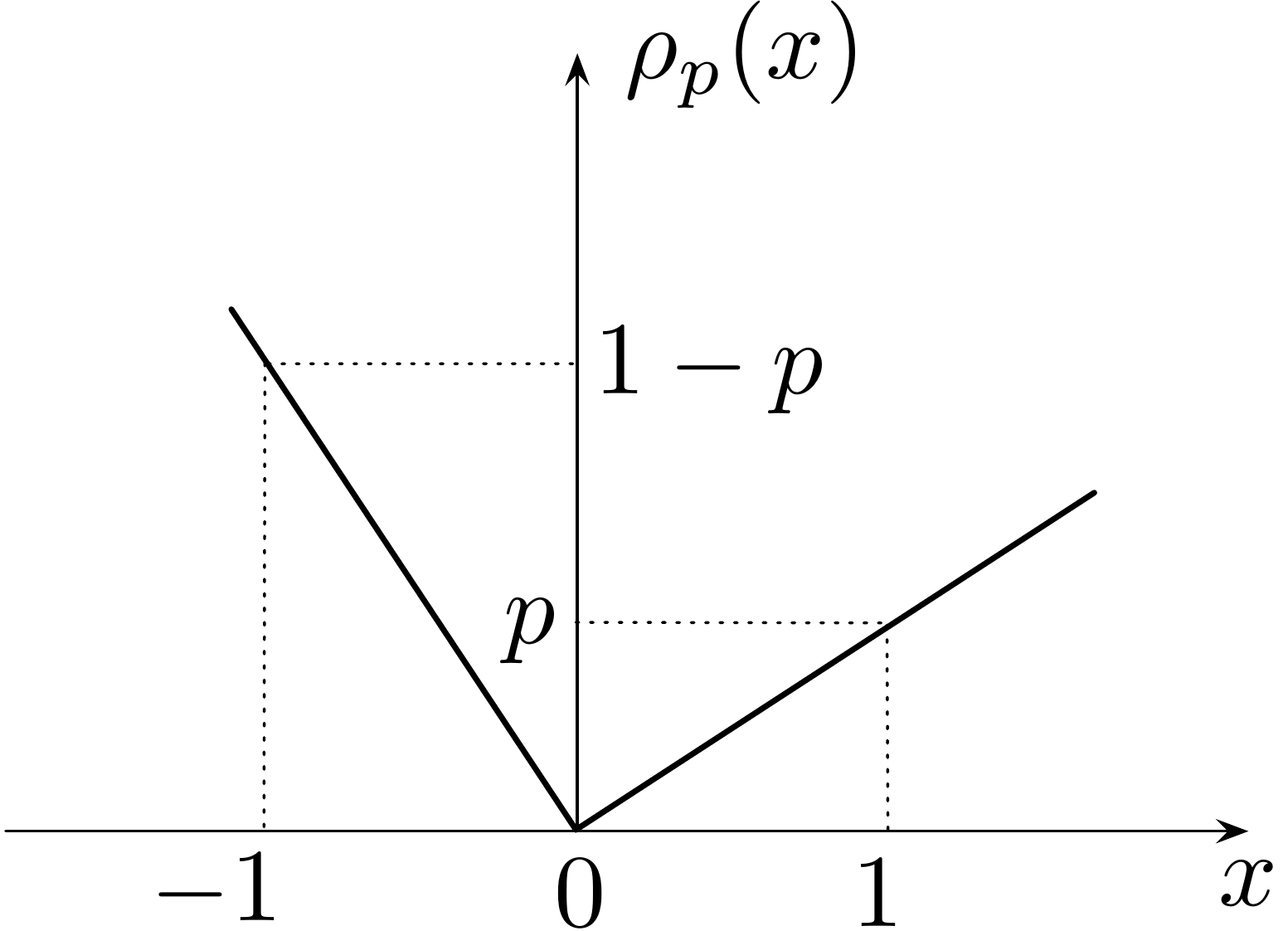}
 \caption{Score function} 
\label{fig:score}
\end{wrapfigure}

Let the local (private) functions be defined as
$ f_i(\xi)\Equaldef \rho_p(z_i-\xi),$, which are called the score functions (see \cref{fig:score}),
and the objective be defined as the aggregate score function
$ f(\xi)\Equaldef\sum_{i=1}^{n}f_i(\xi),
$
then the sample quantile is the solution of the following distributed optimization problem:
\begin{equation}\label{eq:main_problem}
\theta_p = \arg \min_{\xi\in \mathbb{R}} \ \ f(\xi)=\sum_{i=1}^{n}f_i(\xi).
\end{equation}

A few noteworthy aspects of \cref{eq:main_problem} are: (1) This is a convex optimization problem;
(2) The objective function is nonsmooth; (3) The local functions have bounded subgradients:
\begin{equation}
|g_i(\xi)| \leq \max\{p,1-p\} \leq 1, \ \ g_i \in \partial f_i;
\end{equation}
and (4) the $p$-quantile $\theta_p$ belongs to the dataset $\mathcal{D}$, for any parameter $p \in \mathcal{P}$, where
    \begin{equation}\label{eq:set_of_p's}
         \mathcal{P} \Equaldef (0,1)\backslash\left\{ \frac{1}{n},\cdots,\frac{n-1}{n}\right\}.
    \end{equation}

This framework can be used to compute many statistics of interest. For example, to compute the maximum $(k=1)$, let $p\in (1-1/n,1)$. To compute the minimum ($k=n$), let $p\in (0,1/n)$. Provided the number of samples in $\mathcal{D}$ is odd, to compute the median, set $p\in \big((n-1)/2n,(n+1)/2n\big)$. In general, if we would like to find the $k$-th largest element of $\mathcal{D}$, then 
\begin{equation}
p \in \left(\frac{n-k}{n},\frac{n-k+1}{n} \right).
\end{equation}

\section{Distributed algorithm and main result}

Herein, we propose and analyze the convergence of a two-time scale distributed sample quantile estimation algorithm in the presence of noise in the communication links. In particular, given the number of sensors $n$ and any probability $p \in\mathcal{P}$, we prove that the algorithm converges to the sample quantile $\theta_p$. The non-smoothness of the empirical cummulative distribution function leads to oscillation around the optimal solution in the process of convergence, which is a difficulty that must be addressed. A novel analysis technique is introduced to tackle this problem.  

Consider a connected undirected graph $\mathcal{G}=([n],\mathcal{E})$ with $n$ nodes, each node observing a real number $z_i$, $i\in[n]$. Here, $[n]=\{1, \ldots, n\}$ denotes the set of nodes and $\mathcal{E} \subset [n] \times [n]$ denotes the set of edges between nodes. 
Let $\mathcal{N}_i$ denote the set of neighbors of the $i$-th node, and $d_{\max}=\max_{i} |\mathcal{N}_i|$. 
Let $L$ denote the graph Laplacian matrix and $\lambda_1,\lambda_2,\cdots,\lambda_n$ denote the eigenvalues of $L$, which satisfy $0=\lambda_1<\lambda_2<\cdots<\lambda_n$. Let $w_i(t)$ denote the local estimate of $\theta_p$ computed by node $i$ at the $t$-th iteration, $m_i(t)$ denotes the message sent by node $i$ to its neighbors $\mathcal{N}_i$, and $v_{ij}(t)$ be the communication noise on the link between nodes $i$ and $j \in \mathcal{N}_i$.

\begin{assumption}\label{assumption}
We assume that the noise sequences of random variables $\{v_{ij}(t)\}_{t=0}^{n}$ are independent and identically distributed (i.i.d.) satisfying the following two properties:
\begin{equation} \label{assump: E_v}
    \mathbf{E}\big[v_{ij}(t)\big] = 0   ~~ \text{and}~~
    \mathbf{E}\big[v^2_{ij}(t)\big] < \infty, \ \ (i,j)\in \mathcal{E}. 
\end{equation}
\end{assumption}

Let $\alpha(t)$ and $\beta(t)$ be two deterministic step-size sequences. We initialize $w_i(0)=z_i, \ \  i\in [n].$ On the $t$-th round of local communication we perform the following steps:

\begin{enumerate}

\item \textbf{Message computation:}
    \begin{equation} 
        m_i(t)=w_i(t)-\alpha(t) \big(\mathbf{1}\big(w_i(t)\ge z_i\big)-p\big).
    \end{equation}

\item \textbf{Local estimate update:}
    \begin{equation}
  w_i(t+1) = m_i(t) - \beta(t)\sum_{j\in \mathcal{N}_{i}}\big(m_i(t)-y_j(t)\big),
   \end{equation}
    where $y_j(t) = m_j(t)+v_{ji}(t)$.

\end{enumerate}

\begin{theorem} \label{thm: QuantileCovergence}
For a given set of samples $\mathcal{D}=\{z_i\}_{i=1}^n$ distributed over $n$ agents connected by a static undirected network $\mathcal{G}$, and 
any $p\in\mathcal{P}$, there exist step-size sequences $\alpha(t)$ and $\beta(t)$ satisfying  
\begin{align} \label{ineq: finite_sum_aa}
        \sum_{t=0}^{\infty} \alpha(t) &=\infty, ~~ \sum_{t=0}^{\infty} \alpha^2(t)<\infty,\\
 \label{ineq: finite_sum_bb}
    \sum_{t=0}^{\infty} \beta(t)&=\infty, ~~ \sum_{t=0}^{\infty} \beta^2(t)<\infty,\\
 \label{ineq: finite_sum_aab}
&   \sum_{t=0}^{\infty} \frac{\alpha^2(t)}{\beta(t)}<\infty,
\end{align} 
 such that the sequence $w_i(t)$ computed acording to \cref{alg: distributed_quantile_estimation} satisfies
   $w_i(t) \overset{\textup{a.s.\ }}{\longrightarrow}\theta_p, \ \ i \in [n]$.
\end{theorem}

\begin{algorithm}[t!] 
	\SetAlgoLined 
	\caption{Distributed two-time scale sample quantile estimation under communication noise}\label{alg: distributed_quantile_estimation}
	\KwIn{measurements $\{z_i\}_{i=1}^n$, initialization $w_i(0)=z_i$, $i\in [n]$, two step-size sequences  $\{\alpha(t)\},\{\beta(t)\}$, quantile parameter $p\in(0,1)$}
	\For{$t=0,1,\cdots$}{
		
		Local subgradient step: 
		$m_i(t)=w_i(t)-\alpha(t) \big(\mathbf{1}\left(w_i(t)\ge z_i\right)-p\big)$

        Send message: $m_i(t)$ to neighbors $j\in \mathcal{N}_i$

        Receive messages: $y_j(t)=m_j(t)+v_{ji}(t)$, $j\in \mathcal{N}_i$
		
		Estimate update step:
			$w_i(t+1) = m_i(t) - \beta(t)\sum_{j\in \mathcal{N}_{i}}\big(m_i(t)-y_j(t)\big)$
	}
	%
\end{algorithm}

\section{Almost sure convergence analysis}

Our proof relies on a two-time-scale stochastic approximation algorithm. Define the deterministic step-size sequences 
 \begin{equation}
 \alpha(t)\Equaldef \frac{\alpha_{0}}{(t+1)^{\tau_1}} \ \ \text{and} \ \ \beta(t)\Equaldef \frac{\beta_{0}}{(t+1)^{\tau_2}},
 \ \ \text{where} \ \
\beta_{0} \le \frac{2}{\lambda_2+\lambda_n},\ \ \textcolor{black}{\alpha_0 \geq 1,}
\end{equation}
and $\tau_1,\tau_2$ satisfy $
0.5< \tau_2 < \tau_1 \le 1$ and $2\tau_1-\tau_2>1$.

\cref{alg: distributed_quantile_estimation} can be expressed in vector form as
\begin{align} \label{alg: quantile_vector}
w(t+1) &\Equaldef  \big(I - \beta(t)L\big)\big(w(t)- \alpha(t)g(t)\big) +\beta(t)v(t).
\end{align}
where 
\begin{multline}
w(t)\Equaldef \big[w_1(t),\cdots,w_n(t)\big]^T, \ \ v(t)\Equaldef\Big[\sum_{j\in \mathcal{N}_{1}}v_{1j}(t),\cdots,\sum_{j\in \mathcal{N}_{n}}v_{nj}(t)\Big]^T, \\ g_i(t) \Equaldef \mathbf{1} \big(w_i(t) \ge z_i\big) - p, \ \ \text{and} \ \ g(t)\Equaldef\big[g_1(t),\cdots,g_n(t)\big]^T.
\end{multline}
Finally, define the following averages 
    \begin{align*}
   \bar{w}(t)\Equaldef\frac{1}{n}\sum_{i=1}^n w_i(t), \ \ 
   \bar{v}(t)\Equaldef\frac{1}{n}\sum_{i=1}^n v_i(t),  \ \ \text{and} \ \ 
\bar{g}(t)\Equaldef\frac{1}{n}\sum_{i=1}^n g_i(t).
   \end{align*}

Next, we prove the almost surely convergence of $w_i(t)$  to the quantile $\theta_p$ defined in the dynamical system in \cref{alg: quantile_vector}, where $p\in \mathcal{P}.$

\vspace{5pt}

\begin{proof}
Start by rewriting $w_i(t) \overset{\textup{a.s.\ }}{\longrightarrow}\theta_p, \ \ i \in [n]$ as
\begin{equation} \label{eq: QuantileCovergence_2}
	\lim_{t \to +\infty} \|w(t) - \theta_p 1\|=0 ~~ \textup{a.s.}
\end{equation}
From the triangle inequality, we have
$
\|w(t) - {\theta_p}1 \| 
\le \|w(t) - \bar{w}(t)1 \| +\| \bar{w}(t)1-{\theta_p}1\|.
$
We will show Eq. \eqref{eq: QuantileCovergence_2} holds by first proving that 
\begin{equation}
\label{eq: converge_to mean}
	\lim_{t \to +\infty} \|w(t) - \bar{w}(t)1 \|=0 ~~ \textup{a.s.}, \ \ \text{and} \ \ 
\lim_{t \to +\infty} \| \bar{w}(t)-{\theta_p}\| = 0 ~~ \textup{a.s.} 
\end{equation}

\vspace{5pt}

%

   
\noindent{\textbf{Part 1:}} 
  \   Recall that the Laplacian matrix $L$ satisfies $L 1 = 0$. Multiplying both sides of \cref{alg: quantile_vector} by $G\Equaldef \frac{1}{n} 1 1^T$  yields
    \begin{equation} \label{eq: secondterm}
    	\bar{w} (t+1) 1  =  G \big(w(t)- \alpha(t)g(t)\big)+ \beta(t) Gv(t).
    \end{equation}
    Let $B(s)\Equaldef\big(I - \beta(s)L-G\big)$. Since $B(s) 1=0$, by combining Eqs. \eqref{alg: quantile_vector} and \eqref{eq: secondterm} we get
    \begin{equation} \label{eq:thetap_unbiasedness}
    w(t+1)-\bar{w} (t+1) 1
     	=
    B(t)\big(w(t)-\bar{w}(t)1- \alpha(t)g(t)\big)+\beta(t)(I-G)v(t). 
\end{equation}
    Therefore, after taking the  norm and squaring both sides of Eq. \eqref{eq:thetap_unbiasedness}, we obtain
    \begin{multline} \label{eq:thetap_unbiasedness_2}
    \norm{w(t+1)-\bar{w} (t+1) 1}^2 
    =
    \norm{B(t)\big(w(t)-\bar{w}(t)1- \alpha(t)g(t)\big)}^2
    +\norm{\beta(t)(I-G)v(t)}^2 \\
    +2\ip{B(t)\big(w(t)-\bar{w}(t)1- \alpha(t)g(t)\big)}{\beta(t)(I-G)v(t)}. 
\end{multline}

    Let $\mathcal{F}_t$ denote the $\sigma$-algebra generated by $\{v(\ell)\}_{\ell=1}^{t-1}$.
    Taking the conditional expectation of \cref{eq:thetap_unbiasedness_2} with respect to $\mathcal{F}_t$, we obtain
     \begin{equation} \label{eq:E_thetap_unbiasedness_1}
    	\E[\norm{w(t+1)-\bar{w} (t+1) 1}^2 \mid \mathcal{F}_t] 
    	=\norm{B(t)\big(w(t)-\bar{w}(t)1- \alpha(t)g(t)\big)}^2 + \E\[\norm{\beta(t)(I-G)v(t)}^2\],
    \end{equation}
	where we used the fact that $\E\big[v(t)\mid\mathcal{F}_t\big]=0$. Let $\eta>0$, using Young's inequality\footnote{$(a+b)^2\le (1+\eta)a^2+(1+1/\eta)b^2, ~\eta>0$.}, we have
	\begin{multline} \label{eq:E_thetap_unbiasedness_2}
    	\E\big[\norm{w(t+1)-\bar{w} (t+1) 1}^2\mid \mathcal{F}_t\big] 
     \leq (1+\eta)\norm{B(t)\big(w(t)-\bar{w}(t)1\big)}^2 \\
    + \(1+1/ \eta\)\norm{\alpha(t)B(t)g(t)}^2+\E\[\norm{\beta(t)(I-G)v(t)}^2\].
    \end{multline}
Using the inequality $\|Ab\| \le \|A\|\|b\|$, we obtain
	\begin{multline} \label{eq:E_thetap_unbiasedness_3}
    	\E\big[\norm{w(t+1)-\bar{w} (t+1) 1}^2\mid \mathcal{F}_t\big] 
    \le(1+\eta)\norm{B(t)}^2\norm{w(t)-\bar{w}(t)1}^2 \\
    +\(1+1/ \eta\)\alpha^2(t)\norm{B(t)}^2\norm{g(t)}^2 +\beta^2(t)\norm{I-G}^2\E\[\norm{v(t)}^2\].
    \end{multline}
We can also show that 
 \begin{equation}\label{eq:norm}
    	\norm{B(s)}=\max\big\{|1-\lambda_2\beta(s)|,|\lambda_n\beta(s)-1|\big\} = 1-\lambda_2\beta(s).
\end{equation}

Incorporating \cref{eq:norm} and $ \norm{I-G}=1$ into Eq. \eqref{eq:E_thetap_unbiasedness_3}, the following inequality holds for any $\eta>0$:
    \begin{multline} \label{eq:E_thetap_unbiasedness_4}
    \E\big[\norm{w(t+1)-\bar{w} (t+1) 1}^2 \mid \mathcal{F}_t\big] 
    \le(1+\eta)\big(1-\lambda_2 \beta(t)\big)^2\norm{w(t)-\bar{w}(t)1}^2 \\
    +\(1+1/\eta\)\alpha^2(t)\big(1-\lambda_2 \beta(t)\big)^2\norm{g(t)}^2
    +\beta^2(t)\E\[\norm{v(t)}^2\].
    \end{multline}
Choosing $\eta=\lambda_2 \beta(t)$, we get 
    \begin{multline} \label{eq:thetap_unbiasedness_6}
    \E\big[\norm{w(t+1)-\bar{w} (t+1) 1}^2 \mid \mathcal{F}_t\big]
    \le  
     \norm{w(t)-\bar{w}(t)1}^2-\lambda_2 \beta(t)\norm{w(t)-\bar{w}(t)1}^2 \\+\frac{\alpha^2(t)}{\lambda_2 \beta(t)}\norm{g(t)}^2+\beta^2(t)\E\[\norm{v(t)}^2\].
    \end{multline}
    The subgradients of the local functions for the distributed quantile computation problem satisfy $\norm{g_i(t)}<1$. Thus, we have $\norm{g(t)}\le \sqrt{n}$.
    Together with Eqs. \eqref{assump: E_v}, \eqref{ineq: finite_sum_bb} and \eqref{ineq: finite_sum_aab}, we obtain 
    \begin{align}
        \E\[\sum_{t=1}^{\infty} \(\frac{\alpha^2(t)}{\lambda_2 \beta(t)}\norm{g(t)}^2+\beta^2(t)\norm{v(t)}^2\)\]< \infty.
    \end{align}
	Therefore, from Lemma \ref{lm: comparison_theorem} (Robbins-Siegmund Theorem \citep{robbins1971convergence}),
    we conclude that $\norm{w(t)-\bar{w} (t) 1}$ converges almost surely to zero, and $\sum_{t=1}^{\infty}\beta(t)\norm{w(t)-\bar{w}(t)1}^2<\infty.$
If $\tau_2\leq 1$, the sequence $\beta(t)$ is not summable. Therefore, 
	\begin{align}
	\lim_{t\to +\infty} \norm{(w(t)-\bar{w}(t)1}=0 ~~ \textup{a.s.}
	\end{align}
\noindent{\textbf{Part 2:}}
    Multiplying both sides of Eq. \eqref{alg: quantile_vector} by $\frac{1}{n} 1^T$ and subtracting ${\theta_p}$, we have
    \begin{align} \label{eq: w_avg_2}
    \bar{w} (t+1)-{\theta_p}  = \bar{w}(t) -{\theta_p} - \alpha(t) \bar{g}(t) + \beta(t) \bar{v}(t).
    \end{align}
Squaring both sides of \cref{eq: w_avg_2} yields
\begin{equation} \label{eq: w_avg_3}
\big|\bar{w} (t+1)-{\theta_p}\big|^2  = \big|\bar{w}(t) -{\theta_p} - \alpha(t) \bar{g}(t)\big|^2 +\beta^2(t) \big|\bar{v}(t)\big|^2 
+ 2\beta(t) \big[\bar{w}(t) -{\theta_p} - \alpha(t) \bar{g}(t)\big]\bar{v}(t).
\end{equation}     
Taking conditional expectation with respect to 	$\mathcal{F}_t$ and using the fact that $\E\big[\bar{v}(t) \mid \mathcal{F}_t\big]=0$, yields
\begin{equation} \label{eq: w_expecatation_1}
\E\big[\big|\bar{w} (t+1)-{\theta_p}\big|^2 \mid \mathcal{F}_t\big]  = \big|\bar{w}(t) -{\theta_p} - \alpha(t) \bar{g}(t)\big|^2 
+\beta^2(t) \E\big[\bar{v}^2(t)\big].
\end{equation}     
Defining $\varphi(t)\Equaldef \big|\bar{w}(t) -{\theta_p} - \alpha(t) \bar{g}(t)\big|^2$, we have
\begin{equation} \label{eq: mt}
	\varphi(t)=\big|\bar{w}(t) -{\theta_p}\big|^2 +\alpha^2(t)\big|\bar{g}(t)\big|^2
	-2\alpha(t) \big(\bar{w}(t) -{\theta_p}\big)\bar{g}(t).
\end{equation}  


For each $g_i(t)$, the following chain of inequalities holds: 
\begin{IEEEeqnarray}{rCl}
	\big(\bar{w}(t) -{\theta_p}\big)g_i(t) 
	&=&\big(w_i(t)-{\theta_p}\big)g_i(t)+\big(\bar{w}(t) -w_i(t)\big)g_i(t)\\
	&\overset{(a)}{\ge}& f_i\big(w_i(t)\big)	-f_i(\theta_p) +\big(\bar{w}(t) -w_i(t)\big)g_i(t)\\
	&\overset{(b)}{=}& f_i\big(w_i(t)\big)-f_i\big(\bar{w}(t)\big)+f_i\big(\bar{w}(t)\big)-f_i(\theta_p) +\big(\bar{w}(t) -w_i(t)\big)g_i(t)\\
	&\overset{(c)}{\ge}& -2\big|\bar{w}(t) -w_i(t)\big|+f_i\big(\bar{w}(t)\big)-f_i(\theta_p),
\end{IEEEeqnarray}
where $(a)$ follows the convexity of $f_i(x)$
, $(b)$ follows from adding and subtracting $f_i\big(\bar{w}(t)\big)$, and $(c)$ follows the $1$-Lipschitz condition that $|g_i(t)|<1$ and $\big|f_i(x)-f_i(y)\big| \le |x-y|$. Therefore,
\begin{equation}
\big(\bar{w}(t) -{\theta_p}\big)\bar{g}(t) 
	= \frac{1}{n} \sum_{i=1}^n \big(\bar{w}(t) -{\theta_p}\big)g_i(t)
	\ge -\frac{2}{n} \sum_{i=1}^n \big|\bar{w}(t) -w_i(t)\big| + f\big(\bar{w}(t)\big)-f(\theta_p), 
\end{equation}
and
\begin{equation}
\varphi(t) \le \big|\bar{w}(t) -{\theta_p}\big|^2- 2\alpha(t) \Big[f\big(\bar{w}(t)\big)-f(\theta_p)\Big] +\frac{4\alpha(t) }{n} \sum_{i=1}^n \big|\bar{w}(t) -w_i(t)\big| +\alpha^2(t)\big|\bar{g}(t)\big|^2.  \label{eq: mt2}
\end{equation}

Incorporating Eq. \eqref{eq: mt2} into Eq. \eqref{eq: w_expecatation_1}, we get
\begin{multline} 
\E\Big[\big|\bar{w} (t+1)-{\theta_p}\big|^2 \ \big| \ \mathcal{F}_t\Big]
 \le \big|\bar{w}(t) -{\theta_p}\big|^2- 2\alpha(t) \big(f(\bar{w}(t)\big)-f\big(\theta_p)\big) \\
 +\frac{4\alpha(t) }{n} \sum_{i=1}^n \big|\bar{w}(t) -w_i(t)\big| +\alpha^2(t)\big|\bar{g}(t)\big|^2 +\beta^2(t) \E\big[\bar{v}^2(t)\big].
\label{eq: w_Expectation2}
\end{multline}
Since $\theta_{p}$ is the minimizer of $f(\xi)$, we have $f\big(\bar{w}(t)\big)-f(\theta_p)\ge 0$. Therefore,
\begin{equation}
\E \bigg[\sum_{t=1}^\infty\alpha^2(t)\big|\bar{g}(t)\big|^2\bigg]< \infty \ \ \ \textup{and} \ \ \ 
\E \bigg[\sum_{t=1}^\infty\beta^2(t) \big|\bar{v}(t)\big|^2\bigg]< \infty.
\end{equation}

To apply Lemma \ref{lm: comparison_theorem}, we must show that 
\begin{equation}
\E\bigg[\sum_{t=0}^\infty\alpha(t+1) \big|\bar{w}(t+1) -w_i(t+1)\big|\bigg]< \infty.
\end{equation}
Since $\big|\bar{w}(s) -w_i(s)\big| \le  \big\|w(s)-\bar{w}(s) 1\big\|$, using \cref{eq:thetap_unbiasedness} the following inequalities hold
\begin{multline}
	\sum_{t=0}^\infty\alpha(t+1) \big|\bar{w}(t+1) -w_i(t+1)\big|
	\le \sum_{t=0}^\infty\alpha(t+1) \big\|w(t+1)-\bar{w}(t+1) 1\big\|\\
	\le \sum_{t=0}^\infty\alpha(t+1)\prod_{s=0}^{t}\big\|B(s)\big\| \big\|w(0)\big\| +\sum_{t=0}^\infty\alpha(t+1)\sum_{l=0}^{t}\alpha(l)\prod_{s=l}^{t}\big\|B(s)\big\|  \big\|g(l)\big\| \\ 
	+\sum_{t=0}^\infty\alpha(t+1)\beta(t)\|I-G\|\big\|v(t)\big\| +\sum_{t=0}^\infty\alpha(t+1) \sum_{l=0}^{t} \beta(l) \prod_{s=l+1}^{t} \big\|B(s)\big\|  \|I-G\|\big\|v(l)\big\|.
\end{multline}
By using Lemma \ref{lm: upper_bound} and $\norm{B(s)}= 1-\beta(s)\lambda_2<\exp\big(-\beta(s)\lambda_2\big)$,
we get 
\begin{align}
\sum_{t=0}^\infty\alpha(t+1)\prod_{s=0}^{t}\big\|B(s)\big\|&<\infty, ~~
\sum_{t=0}^\infty\alpha(t+1)\sum_{l=0}^{t}\alpha(l)\prod_{s=l}^{t}\big\|B(s)\big\| <\infty,\\
\sum_{t=0}^\infty\alpha(t+1) &\sum_{l=0}^{t} \beta(l) \prod_{s=l+1}^{t} \big\|B(s)\big\| <\infty.
\end{align}
From \cref{assumption}, using Jensen's inequality we get  $\E\big[\|v(l)\| \big] < \sqrt{\E\big[\|v(l)\|^2 \big]}<\infty$. The step-size sequences $\alpha(t)$ and $\beta(t)$ satisfy 
$\sum_{t=0}^\infty\alpha(t+1)\beta(t)<\infty.$
Therefore,
\begin{equation}
	\E \bigg[\sum_{t=1}^\infty\alpha(t) \big|\bar{w}(t) -w_i(t)\big|\bigg]< \infty.
\end{equation}

Thus, Lemma \ref{lm: comparison_theorem} implies that $\big|\bar{w}(t) -{\theta_p}\big|$ converges almost surely and 
\begin{equation}
\sum_{t=1}^\infty\alpha(t) \Big(f\big(\bar{w}(t)\big)-f(\theta_p)\Big)< \infty.
\end{equation}

Since $\alpha(t)$ is not summable, we have
$	\liminf_{t \to \infty} f\big(\bar{w}(t)\big)=f(\theta_p)=0$ almost surely. 
Thus, there exists a subsequence $\big\{\bar{w}(t_j)\big\}$ such that
\begin{equation}
	\lim_{j \to \infty} f\big(\bar{w}(t_j)\big)=\liminf_{t \to \infty} f\big(\bar{w}(t)\big)=f(\theta_p)=0~~ \text{a.s.}
\end{equation}
From \cite[Section 1.3, pp. 7--9]{Koenker:2005}, since $np$ is not an integer, $\theta_{p}$ is the unique minimum of \cref{eq:quantile_cvx}. Furthermore, since $f(\xi)$ is continuous, we have $\lim_{j \to \infty} \big|\bar{w}(t_j)-\theta_p\big|=0$ almost surely.
Together with the fact that $|\bar{w}(t) -{\theta_p}|$ converges almost surely, we have 
\begin{equation} \label{neq: convergence}
	\lim_{t \to \infty} \big|\bar{w}(t)-\theta_p\big|=0~~\text{a.s.}
\end{equation}

\begin{figure}
 \centering
 \includegraphics[width=0.75\columnwidth]{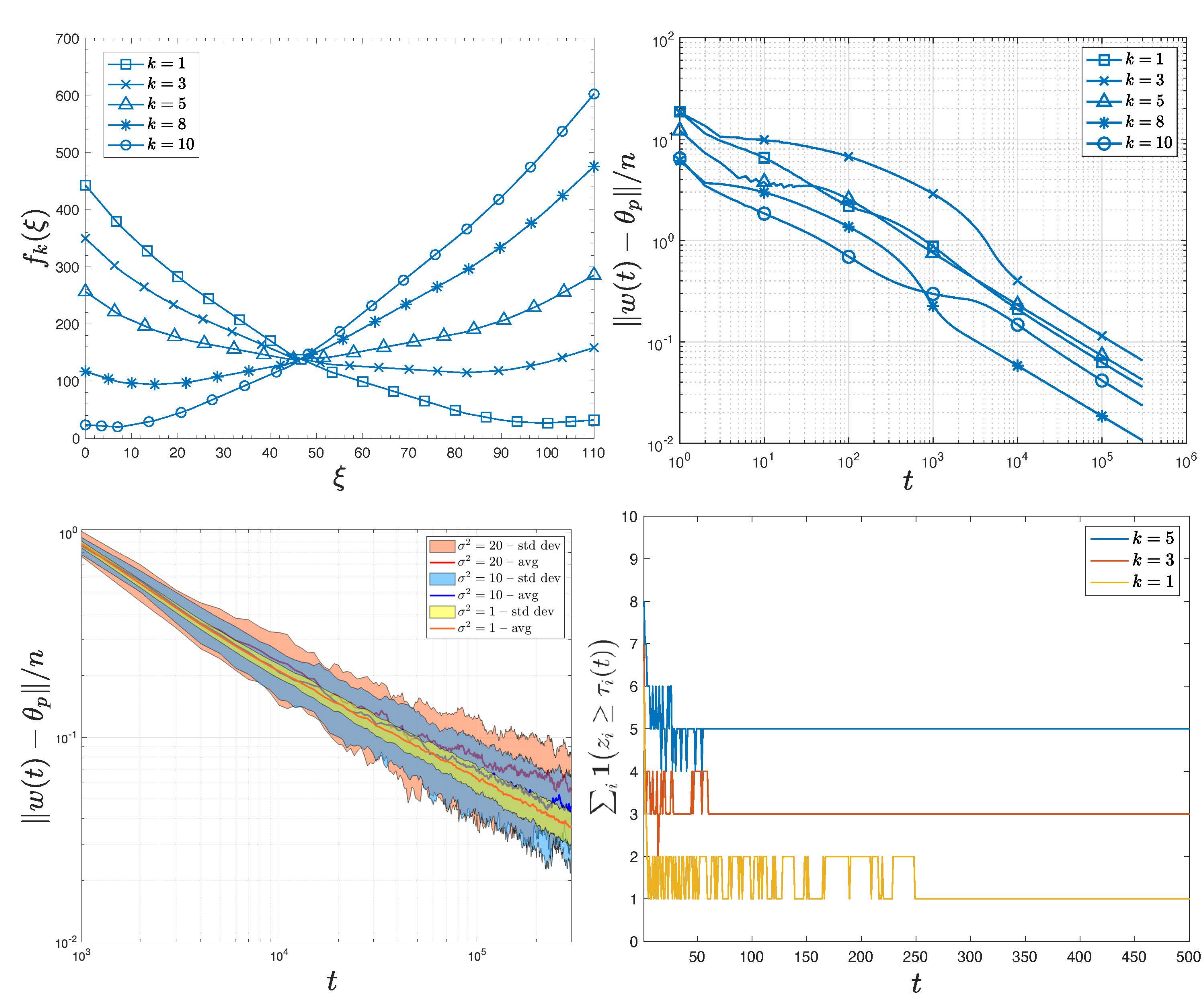}
 \caption{Aggregate score functions (top left) and noiseless distributed sample quantile estimation and  for  different $k$, $k=1,3,5,8,10$ (top right). 
 Convergence of distributed sample quantile inference under different communication noise with $\sigma^2=1,10,20$ and $k=1$ (bottom left). 
 Number of agents with measurements greater than their corresponding thresholds as a function of iterations, for $k=1,3,5$ and $\sigma^2=10$ (bottom right).} 
\label{fig:objective}
\end{figure}

\end{proof}

\vspace{-20pt}

\section{Numerical results}

In this section, we provide some numerical results that display the convergence of the distributed sample quantile inference algorithm and its application to top-$k$ data selection. Consider a system with $n=10$ agents interconnected by the graph in Fig. \ref{fig:system}. The data samples $\mathcal{D}=$ $(45$, $8$, $22$, $91$, $15$, $82$, $53$, $7$, $44$, $99)$ were drawn from an i.i.d. uniform distribution on $\{1,\cdots,100\}$, whose empirical CDF is shown in \cref{fig:ECDF} (left). The communication noises are zero mean i.i.d. Gaussian random variables with variance $\sigma^2$. The parameters of two time-scale sequences are set to $\alpha_0=80$, $\beta_0=2/(\lambda_2+\lambda_n)$, $\tau_1=1$ and $\tau_2=0.505$. 

The panel shown in \cref{fig:objective} displays several aspects that support our theoretical analysis. First, \cref{fig:objective} (top left) shows the aggregate score functions for different values of $k$, and we can see how the curvature varies with the value of $k$ leading to some values of $\theta_p$ being faster to compute than others.
The convergence of distributed sample quantile inference with noiseless links for different $k$ is shown in Fig. \ref{fig:objective} (top right), which shows that the asymptotic  convergence rate (slope of the error curves) are approximately constant. However, the offset among curves depends on $k$, the dataset, and the graph connectivity and is difficult to characterize.

The convergence of distributed sample quantile estimation and top-$k$ selection under noisy links is shown in Fig. \ref{fig:objective} (bottom left). The variance of the noise is set to be $\sigma^2=1, 10, 20$, and $k=1$. We have generated 100 sample paths for each value of $\sigma^2$, and the figure shows the asymptotic convergence of the average of the sample paths. Finally, \cref{fig:objective} (bottom right) shows the convergence of the number of agent with data samples greater than their corresponding thresholds, for $k=1,3,5$ with $\sigma^2=10$. The threshold at each agent is set to be $\tau_i(t)=w_i(t)-0.5$. This correction needs to be done to avoid oscillation of the agent holding the $k$-th largest data sample. 
When the number of iterations is larger than $55$, $61$ and $250$, the system can always find the top-$5$, top-$3$, top-$1$ data points, respectively. Therefore, even though the convergence time in the simulations is in the order of $10^4$ or $10^5$, the convergence to the right decision whether an agent is holding one of the top-$k$ data points happen within a much smaller number of iterations, which is a desirable feature\footnote{All the code used to obtain the numerical results contained herein can be found at \url{https://github.com/mullervasconcelos/L4DC23.git}}
. 

\vspace{-10pt}

\section{Conclusion}
In this work, we have studied networked top-$k$ data selection by using distributed sample quantile inference over noisy channels. We have provided the analysis for the almost sure convergence of a two-time-scale distributed subgradient method. Numerical results have demonstrated the convergence of the distributed algorithm and the corresponding data selection.
There are two interesting directions for future work. One is to analyze the convergence rate for the distributed sample quantile estimation. The other is to provide new algorithms to speed up the convergence.


\vspace{-10pt}

\section{Auxiliary Lemmas}

\begin{lemma}[Theorem 1, \cite{robbins1971convergence}] \label{lm: comparison_theorem}
 Let $\big\{v(t), \mathcal{F}_{t}\big\},\big\{d(t), \mathcal{F}_{t}\big\},$ and $\big\{a(t), \mathcal{F}_{t}\big\}$ be three nonnegative adapted sequences. If 
 \begin{equation}
\mathbf{E}\big[v(t+1) \mid \mathcal{F}_{t}\big] \leq v(t)+a(t)-d(t) \ \ \ \  \text{and} \ \ \ \ \mathbf{E}\bigg[\sum_{t=1}^{\infty} a(t)\bigg]<\infty
 \end{equation}
  then $\sum_{t=1}^{\infty} d(t)<\infty$ almost surely, and $v(t)$ converges
almost surely.
\end{lemma}

\begin{lemma} [Lemma 4.2, \cite{yi2021distributed}] \label{lm: upper_bound} Suppose that  $\alpha(t)$ and $\beta(t)$ satisfy \cref{ineq: finite_sum_aa,ineq: finite_sum_bb,ineq: finite_sum_aab}. Then for any $c>0$, we have
    \begin{equation}
        \sum_{t=0}^\infty\alpha(t+1) \exp\(-c\sum_{s=0}^{t}\beta(s)\)< \infty, \ \ 
\text{and} \ \
        \sum_{t=0}^\infty\alpha(t+1) \sum_{l=0}^{t}\alpha(l) \exp\(-c\sum_{s=l}^{t}\beta(s)\)< \infty.
    \end{equation}
\end{lemma}

\end{document}